\documentclass[prb,superscriptaddress,amsmath,amssymb,reprint]{revtex4-1}

\usepackage{hyperref,xcolor,graphicx,empheq,framed}

\newcommand{\dd}[1]{\mathrm{d}#1\,}

\newcommand{\avg}[1]{\langle{#1}\rangle}
\renewcommand{\Re}{\mathop{\mathrm{Re}}}
\renewcommand{\Im}{\mathop{\mathrm{Im}}}

\DeclareMathOperator{\Tr}{Tr}
\DeclareMathOperator{\tr}{tr}

\begin{document}

\title{Spectral representation of the heat current in a driven Josephson junction}

\author{P.~Virtanen}
\affiliation{NEST, Istituto Nanoscienze-CNR and Scuola Normale Superiore, I-56127 Pisa, Italy}
\author{P.~Solinas}
\affiliation{SPIN-CNR, Via Dodecaneso 33, I-16146 Genova, Italy}
\author{F.~Giazotto}
\affiliation{NEST, Istituto Nanoscienze-CNR and Scuola Normale Superiore, I-56127 Pisa, Italy}

\date{\today}

\begin{abstract}
  We discuss thermal transport through a Josephson junction in a
  time-dependent situation.  We write the spectral representation of
  the heat current pumped by a generic drive.  This enables
  separation of the dissipative and reactive contributions,
  of which the latter does not contribute to long-time averages.  We
  discuss the physical interpretation, and note that the
  condensate heat current identified in [K. Maki and A. Griffin,
    Phys. Rev. Lett. \textbf{15}, 921 (1965)] is purely reactive.  The results
  enable a convenient description of heat exchanges in a Josephson
  system in the presence of an external drive, with possible applications for the
  implementation of new cooling devices.
\end{abstract}

\maketitle

\section{Introduction}

Devices based on quantum mechanical effects could have a huge
technological impact in the next decade.  Quantum computers
\cite{nielsen-chuang}, sensors
\cite{giazotto2010-sqi,giazotto2008-upj,govenius2016-dzm,clarke2006-shb,nagel2011-sqi,vasyukov2013-ssq,ronzani2013-msq},
and metrological devices
\cite{udem2002optical,shapiro1963-jci,tsai1983high,kautz1987precision,giazotto2011-jqe,solinas2015-jrc,pekola2008-hst}
promise to be more efficient, precise and outperform the classical
ones.  However, to work properly they require low and stable working
temperatures.  For this reason, it has become of paramount importance
to be able to manipulate, store and transport energy at the nanoscale
precisely and efficiently.

In this direction, the possibility to coherently control the heat
flowing between two superconductors by manipulating the
superconducting phase difference has attracted much attention
\cite{giazotto2012-jhi,martinez2014-qdt,martinez2015-reh,fornieri2016-npe}.
The main advantage with respect to other nanodevices is that, in some
configurations, the superconducting phase can be controlled directly
through an external magnetic field.  This research field is still
vastly unexplored but could be the ground play for a new class of
quantum devices such as coherent coolers and nano-engines
\cite{solinas2016microwave}.  Yet, to fully understand and exploit the
potentialities of phase-coherent heat control we need to understand
how the energy is transported when the system is subject to a
time-dependent drive.

The dependence of the heat current flowing through a
temperature-biased Josephson junction on the order parameter phase
difference was predicted soon after the discovery of the Josephson
effect, \cite{maki1965,maki1966-eet} but measured only much
later. \cite{giazotto2012-jhi} Several theoretical aspects of the
problem were also clarified only fairly
recently. \cite{frank1997-ecs,guttman1997-pdt,guttman1998-ieh,zhao2003-phase,zhao2004-htt,golubev2013-htt}
Most of the theoretical studies on the Josephson heat transport have
concentrated on steady-state operation, with only few works addressing
the microscopic description of effects from time-dependent
driving. \cite{golubev2013-htt,virtanen2014-ttt}

The heat current through Josephson tunnel junctions was considered for
arbitrary time-dependent phase difference in
Ref.~\onlinecite{golubev2013-htt} based on a BCS tunneling Hamiltonian
calculation, extending results obtained earlier for constant
voltage. \cite{maki1965,maki1966-eet,guttman1997-pdt} Some aspects of
these results appear to be not fully understood, in particular the
interpretation of the ``condensate'' or ``sine'' energy current.
\cite{maki1966-eet,golubev2013-htt} That this current is associated
with the condensate appears clear from the structure of the tunneling
calculation, but its exact interpretation is less clear, given that it
remains nonzero and can have either sign also at $T=0$.  Moreover,
although its contribution to steady-state quantities vanishes in the
cases considered, it is not immediately obvious whether it in general
could contribute to time-averaged quantities in other situations.

In this work, we revisit the previous results.  We write the currents
in a spectral representation, and define associated causal response
functions, which clarifies the general structure.
From this approach, it follows that the condensate component
persisting at $T=0$ is purely \emph{reactive}, and does not contribute
to long-time averages of heat currents, for any form of drive. We
discuss the analytic properties of the reactive components, and point
out a ``quasiparticle'' part not explicitly discussed in previous
works.  Finally, we obtain a simple result for the heat current driven
by an arbitrary periodic drive, and discuss issues relevant to
practical implementation and physical interpretation of the results.

\section{Model}

\begin{figure}
  \includegraphics{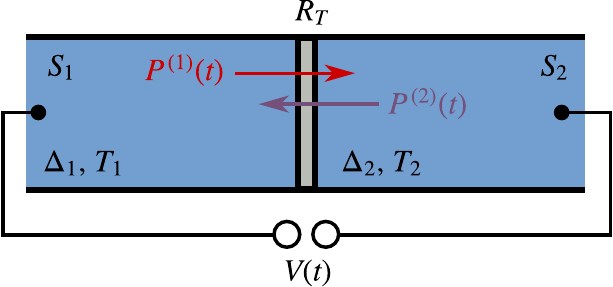}
  \caption{\label{fig:setup} Schematic setup of two superconductors
    $S_1$, $S_2$ with energy gaps $\Delta_1$, $\Delta_2$,
    separated by a tunnel barrier with resistance $R_T$.
    Differences in temperatures $T_1$, $T_2$ and external bias $V(t)$
    drive the heat currents $P^{(1)}(t)$, $P^{(2)}(t)$ between the two superconducting electrodes.
  }
\end{figure}

We consider two superconductors $S_1$ and $S_2$ with superconducting
gaps $\Delta_1$ and $\Delta_2$, respectively, connected by a tunnel
junction of resistance $R_T$.  The superconducting leads are assumed
to be at temperatures $T_1$ and $T_2$ (see Fig. \ref{fig:setup}).  We
consider the corresponding BCS tunneling Hamiltonian model,
\begin{align}
  \label{eq:bcs}
  H &= H_1 + H_2 + H_T
  \,,
  \\
  H_1 &=
  \sum_{k\sigma}[
    \xi_{1k}c_{1k\sigma}^\dagger c_{1k\sigma}
    + (\Delta_1 c_{1k\sigma}^\dagger{}c_{1,-k,-\sigma}^\dagger + \mathit{h.c.})
  ]
  \\
  H_2 &= 
  \sum_{k\sigma}[
    \xi_{2k}c_{2k\sigma}^\dagger c_{2k\sigma}
    + (\Delta_2 c_{2k\sigma}^\dagger{}c_{2,-k,-\sigma}^\dagger + \mathit{h.c.})
  ]
  \\
  H_T &=
  \sum_{kq\sigma}
  e^{i\varphi(t)/2}M_{kq\sigma}c_{1k\sigma}^\dagger{}c_{2q\sigma} + \mathit{h.c.}
  \,.
\end{align}
The time-dependent phase difference $\varphi(t)$ is gauged to the
tunneling Hamiltonian, so that the order parameters $\Delta_1$,
$\Delta_2$ are real-valued. Moreover, a standard unitary
transformation \cite{mahan1993-mp} has been made, shifting energies
relative to the chemical potential, $\xi_k=\epsilon_k-\mu$.

Before starting, it is useful to clarify what we mean by heat
current.  The observable we are interested in is the variation of the
energy of superconductor $i$ in time.  Following previous works
\cite{maki1965,
  guttman1997-pdt,guttman1998-ieh,zhao2003-phase,zhao2004-htt,
  golubev2013-htt} we define the {\it heat current} exiting $S_i$ as
$P^{(i)}\equiv-\frac{d}{dt}\avg{H_i}$.
Notice that despite the fact that this is a well defined observable,
its interpretation in terms of classical thermodynamic quantities,
e.g., in relation to entropy and heat, poses difficulties
\cite{guttman1997-pdt} and will not be discussed here.

Note that as soon as $S_1$ and $S_2$ are coupled through $H_T$, a
fraction of energy is stored as (Josephson) coupling energy.  This
energy is a property of the total system, and cannot be clearly
identified as belonging to either $S_1$ or $S_2$.  At the same time,
the energy flowing out of, say, $S_1$ can either go to $S_2$ or
increase the coupling energy.  This problem is evident in the strong
coupling regime, where the energy associated to $H_T$ can dominate
over the other contributions \cite{campisi2009strong,carrega2016-eed}. If the coupling
energy however is bounded and $S_1$ and $S_2$ are thermodynamically large, the
long-time averages $\overline{P^{(i)}}$ of energy flows can be
expected to be dominated by heat flow to the bulk of the
terminals. \cite{campisi2011-cqf,esposito2010-epa}

After this necessary clarification, we can discuss some general
properties of the energy exchanges that occur between the
superconductors. The rate of change of the total
energy of the system is
\begin{align}
  \dot{W}(t)
  &=
  \partial_t\avg{H(t)}
  =
  -
  P^{(1)}(t)
  -
  P^{(2)}(t)
  +
  \partial_t\avg{H_T(t)}
  \notag
  \\
  &=
  \tr[\rho(t)\partial_tH_T(t)]
  \,,
  \label{eq:powersum}
\end{align}
where the ensemble average is $\avg{A}=\Tr[A\rho]$, and $\rho(t)$ is
the density matrix of the total system. Above, unitarity of the time
evolution, $\dot{\rho}=-i[H,\rho]$, was used.  The time variation
$\dot{W}(t)$ of the Hamiltonian is related to the work done on the
system and the power injected in it \cite{campisi2011-cqf,solinas2013}.  We can write
$\partial_t H_T(t)=i[H_T(t),N_1]\frac{1}{2}\partial_t\varphi(t)$ and,
since the electron current operator $I$ and the voltage $V$ are
proportional to $[H_T(t),N_1]$ and $\partial_t\varphi$, respectively,
we obtain the familiar form for the power injected in a electrical
circuit, i.e., $\dot{W}(t)=I(t)V(t)$.  The power injected into the
total system can thus either increase the energies of the
superconductors or change the coupling energy. As expected, in the
results below the coupling energy term does not contribute to
time-averaged heat currents, and in the time average, the total absorbed heat
current $-\overline{P^{(1)}}-\overline{P^{(2)}}$ is equal to the input
power $\overline{\dot{W}}$.

\section{Spectral representation}

The heat current $P^{(1)}(t)=-\frac{d}{dt}\avg{H_1}$ was
calculated to leading order in tunneling in
Ref.~\onlinecite{golubev2013-htt} for general time-dependent
drive. The result reads
\begin{align}
  P^{(1)}(t) &=  P^{(1)}_J(t) + P^{(1)}_{qp}(t)
  \,,
  \nonumber \\
  \label{eq:Pqp-base}
  P^{(1)}_{qp}(t)
  &=
  \frac{-i}{\pi{}R_T}
  \int_{-\infty}^t\dd{t'}
  e^{-\eta{}(t-t')}
  [\dot{W}^{qp}_1(t-t')W^{qp}_2(t-t')
  \nonumber \\\notag&\quad
    + \dot{W}^{qp}_1(t'-t)W^{qp}_2(t'-t)]
  \cos\frac{\varphi(t)-\varphi(t')}{2}
  \nonumber \\
  P^{(1)}_J(t)
  &=
  \frac{-i}{\pi{}R_T}
  \int_{-\infty}^t\dd{t'}
  e^{-\eta{}(t-t')}
  [\dot{W}^J_1(t-t')W^J_2(t-t')
  \nonumber  \\\notag&\quad
    + \dot{W}^J_1(t'-t)W^J_2(t'-t)]
  \cos\frac{\varphi(t)+\varphi(t')}{2}
 \nonumber  \\
  W^J_j(t)
  &=
  \int_{-\infty}^\infty
  \dd{E}
  e^{-iEt}F(E)[1-f_j(E)]
  \,,
\end{align}
where $f_j(E)= \frac{1}{e^{E/T_j} + 1}$,
$F_j(E)=-F_j(-E)=\Re[|\Delta_j|(E^2-|\Delta_j|^2)^{-1/2}]$,
$W^{qp}_j(t)=\dot{W}^J_j(t)/(i\Delta_j)$ and $\eta\to0^+$.
Above, $R_T$ is the tunnel junction resistance, and we set $e=k_B=\hbar=1$.

General properties of the above result can be more clearly seen in the
spectral representation.  Similarly as in standard discussions of the charge
current, we define \cite{werthamer1966-nso}
\begin{align}
  e^{i\varphi(t)/2}
  &=
  \int_{-\infty}^\infty\frac{\dd{\omega}}{2\pi}
  e^{-i\omega t}
  \Phi(\omega)
  \,.
\end{align}
It will also be convenient to consider the Fourier transform of the
heat current,
$P^{(1)}(\omega)=\int_{-\infty}^\infty\dd{t}e^{i\omega{}t}P^{(1)}(t)$.
Long-time averages can be expressed as
$\overline{P}=\lim_{\tau\to\infty}\overline{[P]}_{\tau}$, where
\begin{align}
  \label{eq:long-avg}
  \overline{[P]}_\tau
  &\equiv
  \int_{-\infty}^\infty
  \dd{t'}
  \frac{z(t'/\tau)}{\tau}
  P(t')
  =
  \int_{-\infty}^\infty
  \frac{\dd{\omega_0}}{2\pi}
  \tilde{z}(\tau\omega_0)^*
  P(\omega_0)
  \,,
\end{align}
where $z$ is some real-valued window function normalized to
$\int_{-\infty}^\infty\dd{x}z(x)=1$ and $\tilde{z}$ its Fourier
transform --- for example a Gaussian, $z(x)=e^{-x^2}/\sqrt{\pi}$,
$\tilde{z}(y)=e^{-y^2/4}$.

Using the definition of $\Phi$ in
Eqs.~\eqref{eq:Pqp-base}
and taking the Fourier transform produces:
\begin{align}
  \label{eq:P1qp}
  P_{qp}^{(1)}(\omega_0)
  &=
  \frac{1}{4i}
  \int_{-\infty}^\infty\frac{\dd{\omega_1}}{2\pi}
  [J^{qp}_1(\omega_1) - J^{qp}_1(\omega_1-\omega_0)^*]
    \\\notag&\times
  [
    \Phi(\omega_1)
    \Phi(\omega_1-\omega_0)^*
    +
    \Phi(-\omega_1)^*
    \Phi(\omega_0-\omega_1)
  ]
  \,,
    \\
   \label{eq:P1J}
  P_J^{(1)}(\omega_0)
  &=
  \frac{1}{4i}
  \int_{-\infty}^\infty\frac{\dd{\omega_1}}{2\pi}
  [J^J_{1}(\omega_1) - J^J_1(\omega_1-\omega_0)^*]
    \\\notag&\times
  [
    \Phi(\omega_1)
    \Phi(\omega_0-\omega_1)
    +
    \Phi(-\omega_1)^*
    \Phi(\omega_1-\omega_0)^*
  ]
  \,,
\end{align}
where the $J$ are causal response functions \cite{mahan1993-mp} defined as
\begin{align}
  \label{eq:J-jfunc}
  J_1^{J/qp}(\omega')
  =
  \frac{i}{\pi R}
  \int_{-\infty}^\infty\frac{\dd{E}}{2\pi}
  \frac{w^{J/qp}(E) + w^{J/qp}(-E)}{\omega' - E + i\eta}
  \,,
\end{align}
and $w^{J/qp}(E)$ are the Fourier transforms of $w^{J/qp}(t)=\dot{W}^{J/qp}_1(t)W^{J/qp}_2(t)$.
The response functions have the symmetry $J^{J/qp}(\omega)=-J^{J/qp}(-\omega)^*$.
By using $F_j(E)=-F_j(-E)$, we can write explicitly
\begin{widetext}
  \begin{align}
    \label{eq:wqp}
    w^{qp}(E) + w^{qp}(-E)
    &=
    -2 \pi i \int_{- \infty}^{\infty} dE' E' N_1(E') N_2(E'-E)   [ f_1(E')- f_2(E'-E)]
    \\
    \label{eq:wj}
    w^{J}(E) + w^{J}(-E)
    &=
    2\pi i \int_{- \infty}^{\infty} dE'
    E' F_1(E') F_2(E'-E)   [ f_1(E')- f_2(E'-E)]
    \,,
  \end{align}
\end{widetext}
where $N_j(E)=N_j(-E)=\Re[E(E^2-|\Delta_j|^2)^{-1/2}]$ is the reduced
density of states. The expressions corresponding to $P^{(2)}$ are
obtained by exchanging the labels $1\leftrightarrow2$ in
Eqs.~\eqref{eq:wqp}--\eqref{eq:wj}.

The above result has a linear response theory form, as expected for
computation for the change in operator expectation values in response
to a perturbation.  Dissipation in linear response is associated with
a specific component --- often the imaginary part --- of the response
functions.  In the results here taking the definition in
Eq.~\eqref{eq:J-jfunc}, under quite general conditions (see below), it
is only the imaginary part that contributes to the long time average
of the heat currents.

The imaginary (``dissipative'') parts can be written as
\begin{align}
\label{eq:ImJqp}
  \Im J_1^{qp}(\omega)
  &=
  \frac{1}{R}
  \int_{-\infty}^\infty\dd{E}
  E
  N_1(E)
  N_2(E-\omega)
   \\
  \notag&\qquad\times
  [
  f_1(E)
  -
  f_2(E-\omega)
  ]
  \nonumber \\
  \label{eq:ImJJ}
  \Im J_1^J(\omega)
  &=
  \frac{-1}{R}
  \int_{-\infty}^\infty\dd{E}
  E
  F_1(E)
  F_2(E-\omega)
   \\
  \notag&\qquad\times
  [
  f_1(E)
  -
  f_2(E-\omega)
  ]
  \,.
  \label{eq:ImJ}
\end{align}
The form of the result suggests they are both associated with
\emph{quasiparticle} transport.  In the normal state,
$\Im{}J_1^{J}(\omega)\rvert_N=0$ and
$\Im{}J_1^{qp}(\omega)\rvert_N=-\frac{\omega^2}{2}+\frac{\pi^2}{6}(T_1^2-T_2^2)$.

In contrast to the imaginary part, the real (``reactive'') part of the
response functions gives only nonzero frequency contributions to the
heat current.  The part $\Re{}J^J_1$ corresponds to the ``condensate''
heat current \cite{maki1965,maki1966-eet,guttman1997-pdt}, and is
related to the ``sine'' heat current of
Ref.~\onlinecite{golubev2013-htt} by
$\Re{}J^J_1(\omega)=-P_{\rm{}sin}^{(1)}(\omega)$. The part $\Re{}J^{qp}_1$ was not
discussed in previous works, as it does not contribute in the
constant-voltage case, but for general drive it is nonzero.

Since the response functions are causal, the reactive parts can be
obtained via Kramers-Kronig relations \cite{mahan1993-mp},
\begin{align}
  \Re J_1^{J/qp}(\omega)
  &= \frac{1}{\pi}\int_{-\infty}^\infty\dd{\omega'}
  \frac{\cal P}{\omega' - \omega}\Im{}J_1^{J/qp}(\omega')
  \\
  &=
  \frac{2\omega}{\pi}\int_{0}^\infty\dd{\omega'}
  \frac{\cal P}{(\omega')^2 - \omega^2}\Im{}J_1^{J/qp}(\omega')
  \notag
  \,.
\end{align}
where $\cal P$ denotes the Cauchy principal value.  The part
$\Re{}J_1^{qp}(\omega)$ is formally divergent,
cf.~\onlinecite{werthamer1966-nso} --- the divergence is regularized
by finite bandwidth/momentum dependence of tunneling.  It can also be
regularized by subtracting
$J_1^{qp}(\omega)\mapsto{}J_1^{qp}(\omega)-\alpha_0-\alpha_1\omega$
with suitable real $\alpha_j$ inside the integral: since
\begin{equation}
  \begin{split}
  \int_{-\infty}^\infty\frac{\dd{\omega_1}}{2\pi}
  [
    \Phi(\omega_1)
    \Phi(\omega_1-\omega_0)^*
    +
    \Phi(-\omega_1)^*
    \Phi(\omega_0-\omega_1)
  ]
  \\
  =
  4\pi\delta(\omega_0)
  \,,
  \nonumber
  \end{split}
\end{equation}
the subtraction does not change the result. We can write
\begin{align}
  \Re J_1^{qp}(\omega)
  &=
  \frac{2\omega}{\pi}\int_{0}^\infty\dd{\omega'}
  \frac{\cal P}{(\omega')^2 - \omega^2}
  \\\notag
  &\times
  [\Im{}J_1^{qp}(\omega') - \Im{}J_1^{qp}(0) + \frac{(\omega')^2}{2}]
  \,.
\end{align}
The normal-state result is $\Re{}J_1^{qp}(\omega)\rvert_N=0$.

\begin{figure}
  \includegraphics{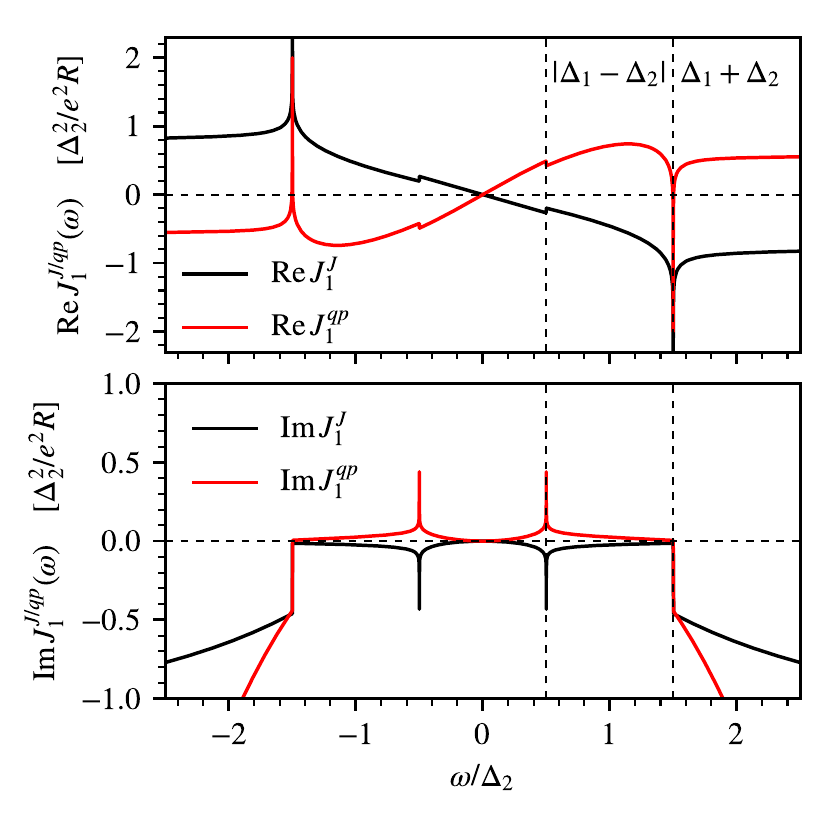}
  \caption{\label{fig:J}
    Real and imaginary parts of the response functions,
    for $\Delta_1=\Delta_2/2$, $T_1=T_2=0.3\Delta_2$.
  }
\end{figure}

Similarly as the charge current response functions,
\cite{werthamer1966-nso} the $J^{J/qp}(\omega)$ functions above have
logarithmic singularities \cite{guttman1997-pdt,zhao2003-phase} that
follow from the gap edge divergences of the BCS density of states. For
$\Im{}J^{J/qp}(\omega)$, the singularities reside at
$\omega=\pm|\Delta_1-\Delta_2|$ and for $\Re{}J^{J/qp}(\omega)$ at
$\omega=\pm|\Delta_1+\Delta_2|$.  By Kramers-Kronig relations, where
$\Im J$ has a discontinuous jump, $\Re{}J$ has a log-singularity, and
vice versa.  If the drive is not resonant, ie., $\Phi(\omega)$ does
not have $\delta$-function or other divergences at exactly these
frequencies, the resulting heat currents remain well-defined.
The response functions are plotted in Fig.~\ref{fig:J}.

Finally, we can comment on the long-time averages.  Based on
Eqs.~\eqref{eq:long-avg},~\eqref{eq:P1qp}, and ~\eqref{eq:P1J}, using
the symmetry of $J^{J/qp}(\omega)$ and $\tilde{z}(y) =
\tilde{z}(-y)^*$, we can write
\begin{align}
  \label{eq:longtime-qp}
  \overline{[P^{(1)}_{qp}]}_\tau
  &=
  \Im\int_{-\infty}^\infty\frac{\dd{\omega}\dd{\omega'}}{8\pi^2}
  \tilde{z}(\tau[\omega-\omega'])^*
  \\\notag&\qquad\times
  [J_1^{qp}(\omega)-J_1^{qp}(\omega')^*]\Phi(\omega)\Phi(\omega')^*
  \,,
  \\
  \label{eq:longtime-J}
  \overline{[P^{(1)}_{J}]}_\tau
  &=
  \Im\int_{-\infty}^\infty\frac{\dd{\omega}\dd{\omega'}}{8\pi^2}
  \tilde{z}(\tau[\omega-\omega'])^*
  \\\notag&\qquad\times
  [J_1^{J}(\omega)-J_1^{J}(\omega')^*]\Phi(\omega)\Phi(-\omega')
  \,.
\end{align}
The average over long time scales $\tau\to\infty$ picks the
zero-frequency component $\omega-\omega'\to0$, and quite generally one
can expand
$\tilde{z}(\tau[\omega-\omega'])^*[J^{J/qp}(\omega)-J^{J/qp}(\omega')^*]\simeq{}\tilde{z}(\tau[\omega-\omega'])^*2i\Im{}J^{J/qp}(\omega)$
inside the integral. As a consequence, only the imaginary parts of the
response functions matter for long-time averages.

\subsection{Sum power}

Consider now the sum power $P^{(T)}=-P^{(1)}-P^{(2)}$. It can be
written in the same form as $P^{(1)}$ in
Eqs.~\eqref{eq:P1qp},\eqref{eq:P1J} but with different response
functions, $J_T^{qp/J}=-J_1^{qp/J}-J_2^{qp/J}$, which can also be
written as
\begin{align}
  \Im J_{T}^{qp/J}(\omega)
  &=
  -\omega
  \Im
  I^{qp/J}(\omega)
  \,,
\end{align}
where \cite{werthamer1966-nso,tucker1985-qda}
\begin{align}
  \Im
  I^{qp}(\omega)
  &=
  \frac{1}{R}
  \int_{-\infty}^\infty\dd{E}
  N_1(E)
  N_2(E-\omega)
  \\
  \notag&\qquad\times
  [
  f_1(E)
  -
  f_2(E-\omega)
  ]
  \\
  \Im
  I^J(\omega)
  &=
  -
  \frac{1}{R}
  \int_{-\infty}^\infty\dd{E}
  F_1(E)
  F_2(E-\omega)
  \\
  \notag&\qquad\times
  [
  f_1(E)
  -
  f_2(E-\omega)
  ]
  \,,
\end{align}
are response functions of the charge current,
\begin{align}
  I(t)
  &=
  -
  \Im
  \int_{-\infty}^\infty\frac{\dd{\omega}}{2\pi}\frac{\dd{\omega'}}{2\pi}
  e^{-i(\omega+\omega')t}
  [\Phi(\omega)\Phi(-\omega')^*I^{qp}(\omega')
  \\\notag&\qquad
  + \Phi(\omega)\Phi(\omega')I^J(\omega')]
  \,.
\end{align}
Noting that
\begin{align}
  \int_{-\infty}^\infty\frac{\dd{\omega}}{2\pi}
  e^{-i\omega t}\Phi(\omega)V(t)
  =
  \int_{-\infty}^\infty\frac{\dd{\omega}}{2\pi}
  (-\omega)e^{-i\omega t}\Phi(\omega)
  \,,
\end{align}
and comparison with
Eqs.~\eqref{eq:long-avg},\eqref{eq:longtime-qp},\eqref{eq:longtime-J} results to
$\overline{I(t)V(t)}=\overline{P^{(T)}}$, i.e.,
$\overline{P^{(1)}}+\overline{P^{(2)}}=-\overline{\dot{W}}$.
There is no average heat current associated with the tunneling energy.

\section{Periodic drive}

Experiments to measure the heat current transferred in superconducting
nanosystems are challenging.  The physical observable is the variation
of temperature of one lead.  Such a measurement is usually done in
the steady state regime when the transient dynamics has vanished.
Under this condition it is natural to assume that the system has a
periodic evolution and study what is the heat current transported in a
period.  Since the Josephson system dynamics is completely
characterized by the superconducting phase, we consider
evolution periodic in the following sense:
\begin{align}
  e^{i\varphi(t+\mathcal{T})/2}
  =
  e^{i\varphi(t)/2}
  \,.
\end{align}
In particular the constant voltage bias discussed in
Refs. \onlinecite{guttman1997-pdt,golubev2013-htt} is periodic in this
sense.  Then,
\begin{align}
  \Phi(\omega)
  &=
  \sum_{k=-\infty}^\infty
  2\pi
  \delta(\omega - \Omega{}k)
  \Phi_k
  \,, \nonumber \\
  \Phi_k
  &=
  \frac{1}{\mathcal{T}}
  \int_{0}^{\mathcal{T}}\dd{t} e^{i\Omega{}kt} e^{i\varphi(t)/2}
  \,,
  \qquad
  \Omega = \frac{2\pi}{\mathcal{T}}
  \,.
\end{align}
Substituting this to Eqs.~\eqref{eq:longtime-qp}--\eqref{eq:longtime-J},
we obtain
\begin{align}
  \label{eq:Pdc-freqresp}
  \overline{P^{(1)}}
  &\equiv
  \overline{P^{(1)}_{qp}}+\overline{P^{(1)}_J}
  =
  \sum_{k=-\infty}^\infty
  |\Phi_{k}|^2
  \Im
  J_1(\Omega k, \phi_k) 
  \,.
\end{align}
where the combined response function appearing above is:
\begin{align}
  \Im
  J_1(\omega,\phi)
  &\equiv
  \Im[J_1^{qp}(\omega) + \cos(\phi) J_1^{J}(\omega)]
  \,.
  \label{eq:ImJ}
\end{align}
and the effective phase difference is
\begin{equation}
  \cos\phi_k
  =
  \frac{2\Re[\Phi_{-k}\Phi_{k}]}{|\Phi_{k}|^2 + |\Phi_{-k}|^2}.
\end{equation}
The long-time average coincides with the average over a single period
--- for $z(x)=\theta(1-x)\theta(x)$ and
$\tau=\mathcal{T}$, $\tilde{z}(\tau\Omega{}[k-k'])^*=\delta_{k,k'}$
in~\eqref{eq:longtime-qp}--\eqref{eq:longtime-J}. 
As above, only the imaginary (``dissipative'') part of the response
function [see Eqs. (\ref{eq:ImJqp}) and (\ref{eq:ImJJ})] contributes
to the heat current, while the real ("reactive") contribution vanishes
in the periodic average.

From a physical point of view, we see that the heat current is
composed by a standard quasi-particle and an "interference"
contribution, similar to the ones in the steady state.
\cite{guttman1997-pdt,zhao2003-phase,giazotto2012-jhi}
Both can be
interpreted \cite{zhao2003-phase} as heat transported by
\emph{quasiparticles}, as can be seen by the presence of the Fermi
function in Eq.~\eqref{eq:ImJ}.

The result in Eq.~\eqref{eq:ImJ} encompasses the ones in the previous
works
\cite{maki1965,guttman1997-pdt,guttman1998-ieh,golubev2013-htt}. For
zero external voltage $\Omega=0$, the result reduces to the expression
for the dc heat tunneling
current. \cite{maki1965,guttman1998-ieh,frank1997-ecs} For constant
voltage, the result recovers that of
Ref.~\onlinecite{golubev2013-htt,guttman1997-pdt}, and for
$\Delta_1=0$ the NIS junction cooling power. \cite{giazotto2006-omi}

\begin{figure}
  \includegraphics{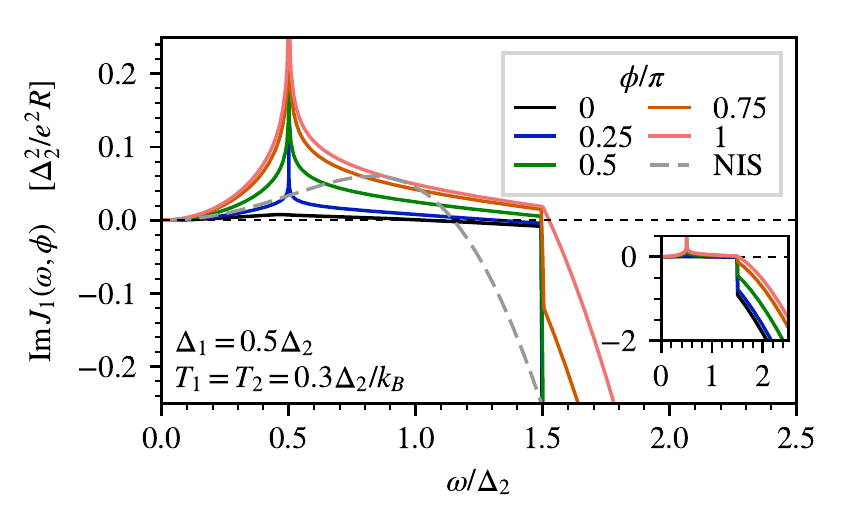}
  \caption{
    \label{fig:R-vs-w}
    Response function $\Im{}J_1(\omega,\phi)$ for
    $\Delta_1=\Delta_1/2$ and $T_1=T_2=0.3\Delta_2/k_B$ for varying $\phi$.
    The NIS case ($\Delta_1=0$) is also shown (dashed line).
    Inset: Same plot with larger y-axis range.
  }
\end{figure}

\begin{figure}
  \includegraphics{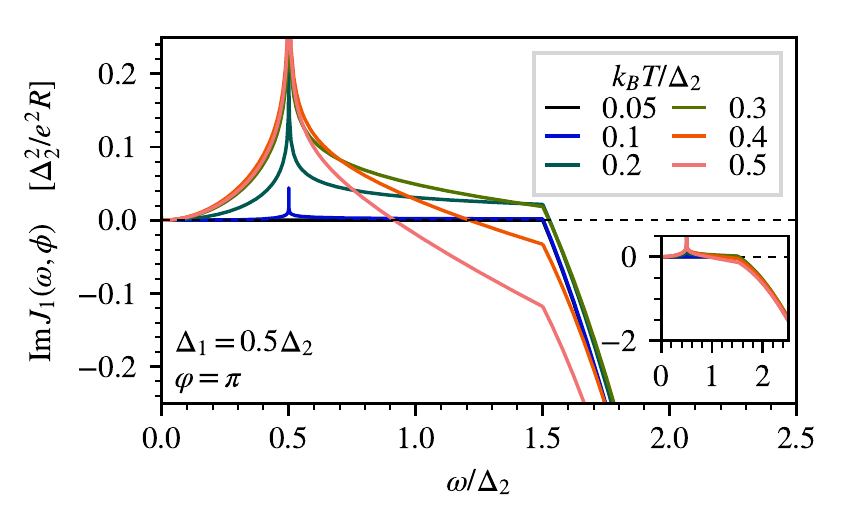}
  \caption{
    \label{fig:R-vs-T}
    Response function $\Im{}J_1(\omega,\phi)$ for
    $\Delta_1=\Delta_1/2$ and varying $T_1=T_2=T$ at $\phi=\pi$.
    Inset: Same plot with larger y-axis range.
  }
\end{figure}

The combined response function is shown in Figs.~\ref{fig:R-vs-w} and
\ref{fig:R-vs-T}.  At low frequencies $\omega<\Delta_1+\Delta_2$ the
function remains positive (cooling), with a logarithmic divergence
appearing at $\omega=|\Delta_1-\Delta_2|$.  As a function of $\phi$,
the maximum is obtained at $\phi=\pi$. At high frequencies
$\omega>\Delta_1+\Delta_2$, quasiparticle transport activates and
leads to a relatively larger but finite negative (heating) result
$J\propto{}-\omega^2$ due to photoassisted pair breaking and
quasiparticle transport.

\section{Discussion and conclusions}

The standard spectral representation expresses clearly the general
properties of the tunneling heat current. Here, it directly indicates
that the reactive ``condensate'' component cannot contribute to
long-time averages. Moreover, a simple result is obtained relating the
dc heat current to the imaginary part of a response function and the
Fourier components of the drive.

The physical interpretation of the results should be viewed in the
context of discussions on heat currents in coupled quantum
systems. \cite{campisi2011-cqf,esposito2010-epa,carrega2016-eed} In
particular, the problem of identifying the coupling energy stored in
the junction raises questions on the status of $P^{(1)}$ defined above
as experimentally relevant observables. While their long-time averages
can be argued to be associated with heat that is accessible to
experiments probing the bulk of the superconducting terminals, what
part of the oscillating components would be accessible by measurements
away from the junction region is not answered by a tunneling
Hamiltonian calculation.  Problems in interpretation are also
illustrated by the zero-temperature behavior: \cite{esposito2015-nhs}
Although for the long-time averages $\overline{P^{(1)}}\le0$ at $T=0$
(only heating is possible at $T=0$), for the instantaneous currents
$P^{(1)}(t)>0$ is possible due to the reactive components that do not
have a definite sign at $T=0$.

We can also note that arguments similar to the above can have
also some implications on the more general discussion on definition of
heat currents in coupled quantum systems,
\cite{ludovico2014-det,esposito2015-nhs,bruch2016-qtd} when the time
dependence is in the coupling Hamiltonian. Based on linear response
theory, time-dependent reactive components are a general feature of
energy currents defined in terms of operator expectation values in
such models.  The Kramers-Kronig relations can then imply constraints
for their time-dependent behavior and interpretation.

In summary, we wrote a spectral representation for the energy current
in Josephson junctions, to obtain a clear picture of the energy
currents predicted in tunnel Hamiltonian calculations.  Being
relatively simple, the results open the way to the
practical design and optimization of SIS' coolers working on pulsed
drive cycles, and for an improved understanding of their general
performance properties.

\section*{Acknowledgments}

We acknowledge S. Gasparinetti for fruitful discussion.   P.S. have
received funding from the European Union FP7/2007-2013 under REA Grant
agreement No. 630925 - COHEAT and the MIUR-FIRB2013 - Project Coca
(Grant No. RBFR1379UX).  P.V. and F.G. acknowledges the European Research
Council under the European Union's Seventh Framework Program
(FP7/2007- 2013)/ERC Grant agreement No. 615187-COMANCHE.

\bibliography{condheat}

\end{document}